\documentclass[aps,prc,showpacs,nofootinbib,twocolumn]{revtex4-1}

\usepackage{epsfig}  
\vspace*{1.0cm}
\usepackage{graphicx,amsmath}
\usepackage{color}
\newcommand\ba{\begin{eqnarray}}
\newcommand\ea{\end{eqnarray}}

\newcommand{\be}{\begin{equation}}
\newcommand{\ee}{\end{equation}}
\newcommand{\bas}{\begin{eqnarray*}}
\newcommand{\eas}{\end{eqnarray*}}
\begin{document}
\vspace*{-1.5cm}
\title{Systematics of  oscillatory behavior in hadronic masses and widths}

\author{Boris Tatischeff}
\email{tati@ipno.in2p3.fr}
\affiliation{CNRS/IN2P3, Institut de Physique Nucl\'eaire, UMR 8608, Orsay, F-91405\\
 and Univ. Paris-Sud, Orsay, F-91405, France}
\vspace*{1cm}
\begin{abstract}
A systematic study of hadron masses and widths shows regular oscillations that can be fitted by a simple cosine function. This property can be observed when  the difference between adjacent masses of each family is plotted versus the mean mass.
\end{abstract}
\maketitle

\section{Introduction}

The non-perturbative structure of hadrons is the object of intensive experimental and theoretical studies. The origin of the mass itself is not yet elucidated as the constituent quarks bring only few percent of the total mass. It is believed that most of the mass is dynamically generated, involving gluons and meson cloud.  In general one can say that the existence of composite hadrons, results from the addition of several forces, related to strong interaction, that combine in, at least, one attractive and one repulsive force. The equilibrium among these forces allows the hadron to exist, otherwise the composite mass will either disintegrate, or mix into a totally new object with loss of the individual components. 

A known example is the charmonium spectrum.  The charmonium is considered the analogue in QCD as positronium in QED. The spectrum of charmonium, like bottomonium,
is very similar (apart from the scale) to the spectrum of positronium. It may be described by a Coulomb-like potential
and a term which takes care of confinement $V\simeq \alpha r+1/2kr^2$. It is understood that the interplay created by repulsive and attractive forces gives rise at the equilibrium to a local harmonic oscillator-like potential. It is therefore not surprising  that static and dynamic properties of hadrons or resonances, as masses, energies and widths show oscillatory behavior, that may be experimentally observable. 
 
Since a rather long time, oscillatory properties have been studied concerning neutrino difference square masses, or particle-antiparticle oscillations.  Structures in time-like proton form factor, have also been recently pointed out \cite{bianconi}. \\

The content of the present work is different.  We analyze bare data of particle masses and widths and  study the mass variation versus the mass increase for adjacent meson and baryon masses of a family (nucleon, pions...and their excitations). We calculate the variation of the function:
\be 
 m_{(n+1)} - m_{n} = f [(m_{(n+1)} + m_{n})/2]\\
\ee
where  $m_{(n+1)}$ corresponds to the (n+1) hadron mass value. The difference of two successive masses is plotted versus the mean  value of the two nearby masses. Such studies can only be illustrated for families holding several, at least five, masses. The result of such variation is shown in a selection of a few figures concerning mesons and baryons. Even though there is no similar equation (like Schrodinger) connecting the masses and widths, the widths are also plotted in order to look at possible regular oscillations.

The obtained data are fitted using a cosine function:
\be
\Delta M = \alpha_{0} + \alpha_{1} cos( (M - M_{0}) / M_{1}) \\
\ee
where M$_{0}$ /M$_{1}$ is defined within 2$\pi$. All coefficients, and masses used to draw the figures are in MeV units. The quantitative informations are given in Table I presented below. The oscillation periods are P = 2 $\pi  M_{1}$. Both $\alpha_{0}$ and $\alpha_{1}$ are adjusted on the extreme values on all figures.

Whereas smaller periods than those given in the table may also reproduce the data, we show in the following figures, the largest possible values. The masses and widths are read from the Review of Particle Physics \cite{pdg}, taking into account all the data reported, even if, in some cases, omitted from the summary table. 

A similar  study was given in Figs. 6 and 7 of a previous paper \cite{chroma}, and, in a work studying the relation between  mass ratios and fractal properties  \cite{fractales}.\\
\begin{figure}[h]
\hspace*{-0.2cm}
\scalebox{0.58}[0.58]{
\includegraphics[bb=29 139  533 544,clip,scale=0.8] {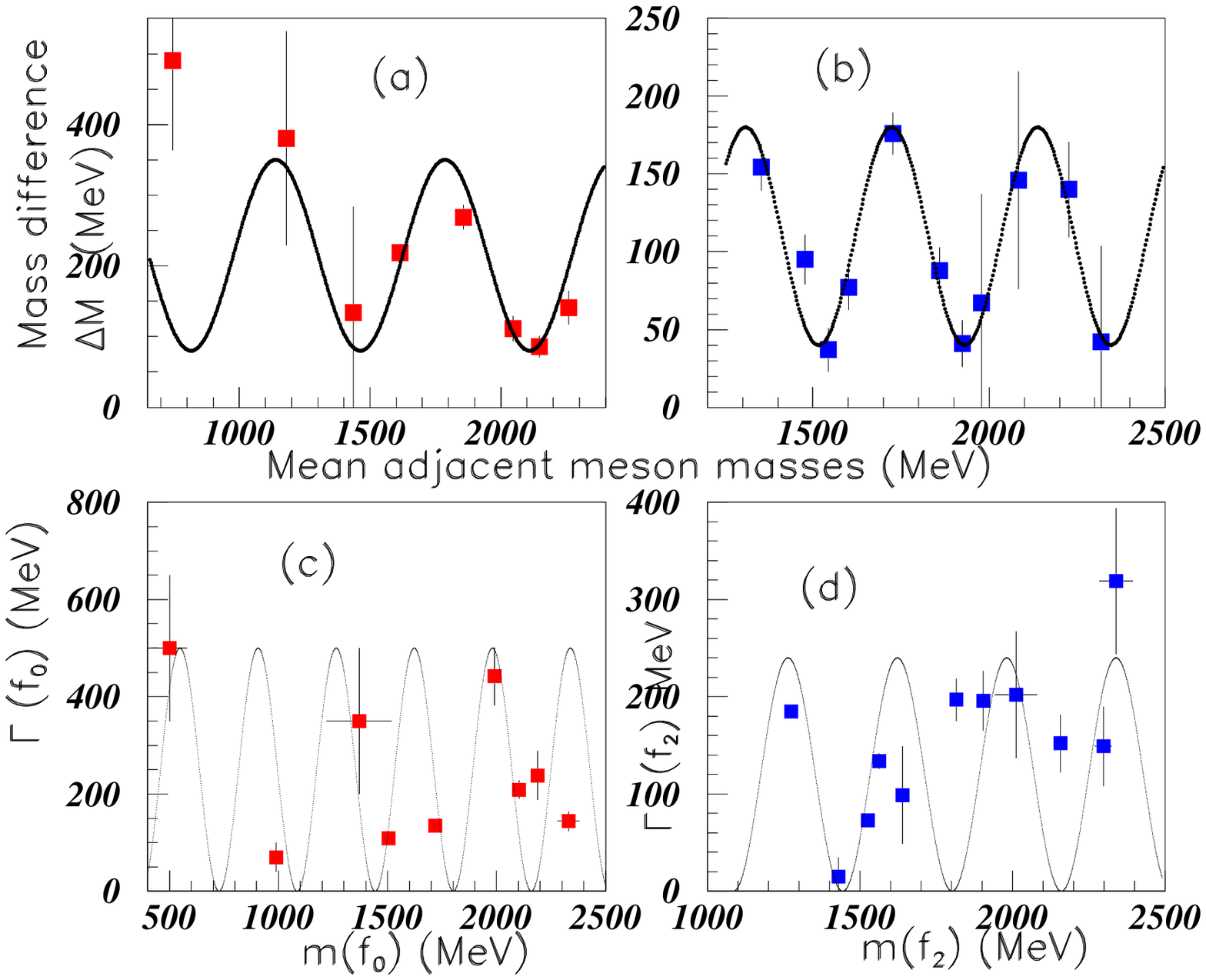}}
\caption{Color on line. Inserts (a) and  (b) show successively the mass difference between successive masses, plotted versus both corresponding mean masses for $f_{0}$ and $f_{2}$ light unflavoured mesons. Inserts (c) and (d) show the  total widths versus the corresponding masses of the same mesons.}
\end{figure}
Fig. 1 shows the results for  $f_{0}$ and $f_{2}$ light unflavoured mesons. Inserts (a) and  (b) show successively the mass difference between successive masses, plotted versus both corresponding mean masses. Inserts (c) and (d) show the  total widths versus the corresponding masses. The $\sigma$ or $f_{0}$(500) meson is broad and its mass (taken to $\Delta$m = 125~MeV), is badly determined. The reasonable fit allows to extrapolate to the masses of the next $f_{0}$ not extracted experimentally up to now.  They are  M $\approx$ 2670 and 2760~ MeV. In the same way, the masses of the next $f_{2}$ mesons can be tentatively predicted to be: M $\approx$ 2380, 2450, and 2625 MeV. 
\begin{figure}[h]
\hspace*{-0.2cm}
\scalebox{0.58}[0.58]{
\includegraphics[bb=20 131 521 551,clip,scale=0.8] {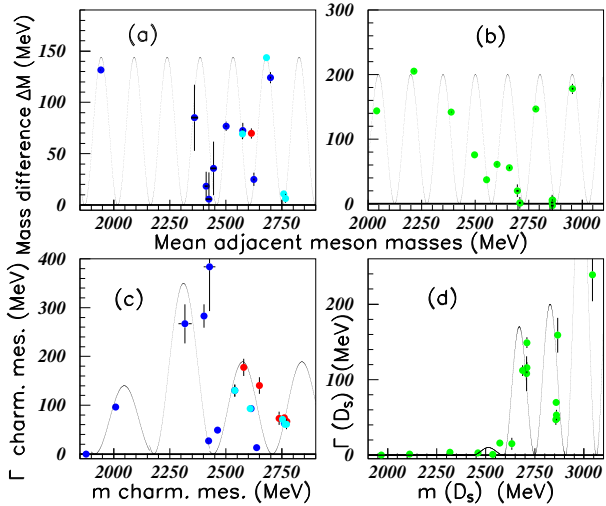}}
\caption{Color on line. Inserts (a) and  (b) show successively the mass difference between successive masses, plotted versus both corresponding mean masses for charmed and charmed, strange mesons. Inserts (c) and (d) show the  total widths versus the corresponding masses of the same mesons.}
\end{figure}

 Fig. 2 shows the results  for charmed and charmed, strange mesons. Mesons with all quantum numbers are kept, otherwise the number of data is too small for given quantum numbers, to get well defined distributions. In insert (a)  the mass difference between successive charmed meson
 masses, plotted versus both corresponding mean masses are shown; Pdg \cite{pdg} data are in blue, LHCB data \cite{batra} are in red and Babar data \cite{babar1} are in sky blue. Many data are overlaided by data from other origin. Insert (c) shows the corresponding data for charmed meson widths plotted versus meson masses.  The corresponding data for charmed strange mesons \cite{qin} are shown in inserts (b) and (d) in green. The periods of oscillation are given in table I.
\begin{figure}[h]
\hspace*{-0.2cm}
\scalebox{0.58}[0.58]{
\includegraphics[bb=21 138  527 550,clip,scale=0.8] {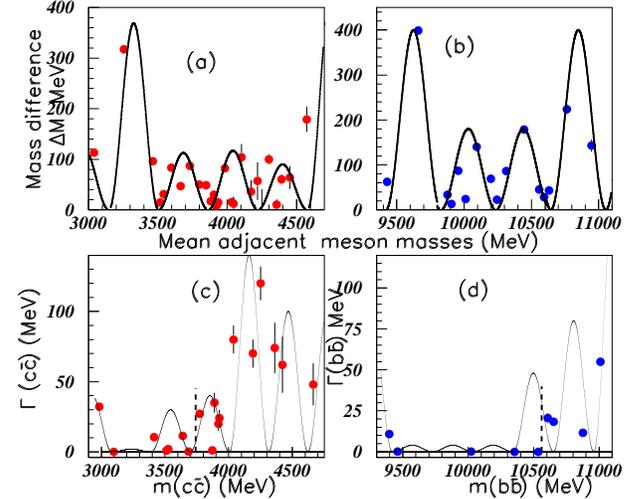}}
\caption{Color on line. Inserts (a) and  (b) show successively the mass difference between successive masses, plotted versus both corresponding mean masses for charmonium (${\it c}{\bar c})$ and  bottomonium (${\it b}{\bar b}$) mesons. Inserts (c) and (d) show the total widths versus the corresponding masses of the same mesons.}
\end{figure} 

 Fig. 3 shows the results for all excited state masses of charmonium (${\it c}{\bar c})$ and  bottomonium (${\it b}{\bar b}$) mesons \cite{pdg}. The shapes of the mass distributions (inserts (a) and (b)) are not very different but their periods differ by a ratio 0.88 (see Table I).
 Both periods of the width distributions are equal: P = 308~MeV.  Since a rather large number of excited masses are known for charmonium and bottomonium mesons, it is possible to compare the previous distributions with the distributions of data selected with given quantum numbers. 
\begin{figure}[h]
\hspace*{-0.2cm}
\scalebox{0.58}[0.58]{
\includegraphics[bb=21 138  527 550,clip,scale=0.8] {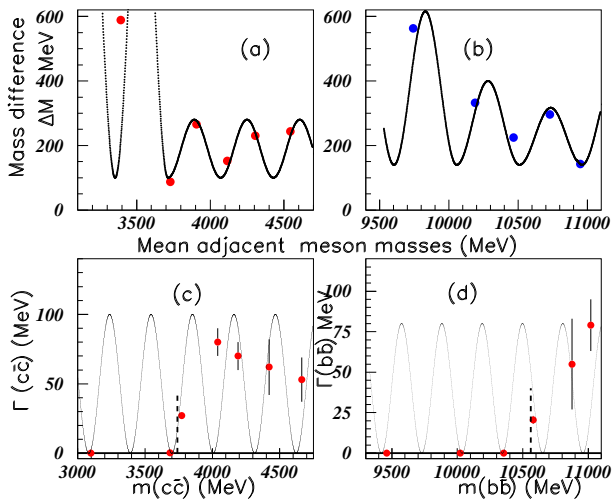}}
\caption{Color on line. Inserts (a) and  (b) show successively the mass difference between successive masses, plotted versus both corresponding mean masses for $0^{-}(1^{--})$ charmonium (${\it c}{\bar c})$ and $0^{-}(1^{--})$ bottomonium (${\it b}{\bar b}$) mesons. Inserts (c) and (d) show the total widths versus the corresponding masses of the same mesons.(See text).}
\end{figure}
\begin{figure}[h]
\hspace*{-0.2cm}
\scalebox{0.58}[0.58]{
\includegraphics[bb=25 133 527 554,clip,scale=0.8] {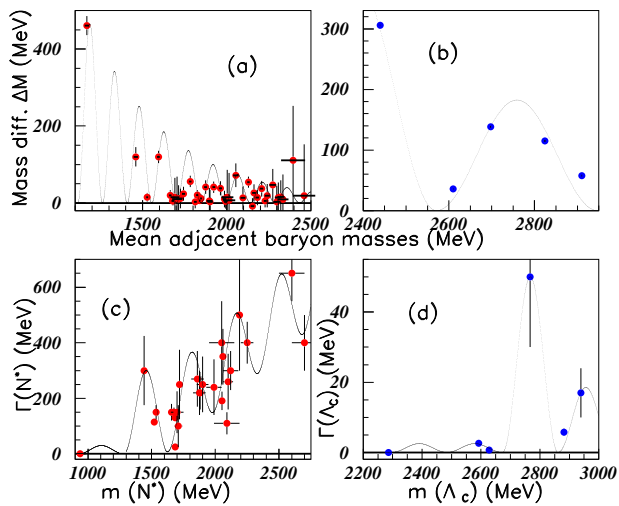}}
\caption{Color on line. Inserts (a) and  (b) show successively the mass difference between successive masses, plotted versus both corresponding mean masses for $N^*$ and $\Lambda_{C}$ baryons.  Inserts (c) and (d) show the  total widths versus the corresponding masses of the same baryons.}
\end{figure}

Fig. 4 shows the results for $0^{-}(1^{--})$ charmonium (${\it c}{\bar c})$ and $0^{-}(1^{--})$ bottomonium (${\it b}{\bar b}$) mesons \cite{pdg}. Not very different shapes are observed for the   charmonium and bottomonium  mass distributions.  The comparison found for the periods of the $0^{-}(1^{--})$ mesons, between the $\Psi$ (${\it c}{\bar c})$ and the $\Upsilon ({\it b}{\bar b}$) is given in table I. They are not modified, except the periods of the bottomonium mass distributions, equal to P = 408~MeV when all quantum numbers are kept, and P = 452~MeV when only $\Upsilon$ mesons are considered. An introduction of new masses may eventually, in the future, reduce this last value. 

The mass of the last quoted meson used in fig. 4(a), X(4660) $?^{?}(1^{--})$ fits perfectly in this  distribution, and is therefore kept,  assigning tentatively the quantum numbers: $I^{G}=0^{-}$.   The extrapolation allows to predict tentatively the next corresponding $\Psi$  masses: M $\approx$ 4805 and 5080~MeV.  In the same way, the tentatively extrapolated $\Upsilon$ masses at the large mass side of insert (b), are: M $\approx$ 11330 and 11560~MeV. In inserts (c) and (d), dashed vertical lignes show the threshold of disintegration into two charm(bottom) mesons, allowing substantially large widths. 

Fig. 5 shows in inserts (a) and (b) the mass difference between successive masses, plotted versus both corresponding mean masses for $N^*$ and $\Lambda_{C}$ baryons.  Inserts (c) and (d) show the  total widths versus the corresponding masses of the same baryons.
When the mass distribution for $N^*$ baryons (insert (a)) displays a rather poor shape, all other three distributions describe the data.
\begin{figure}[h]
\hspace*{-0.2cm}
\scalebox{0.58}[0.58]{
\includegraphics[bb=22 131 533 555,clip,scale=0.8] {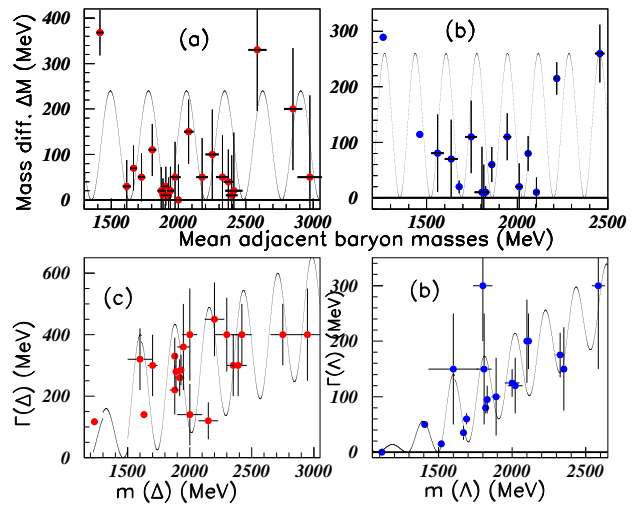}}
\caption{Color on line. Inserts (a) and  (b) show successively the mass difference between successive masses, plotted versus both corresponding mean masses for $\Delta$ and $\Lambda$ baryons.  Inserts (c) and (d) show the  total widths versus the corresponding masses of the same baryons.}
\end{figure}

Fig. 6 shows in inserts (a) and (b) the mass difference between successive masses, plotted versus both corresponding mean masses for $\Delta$ and $\Lambda$ baryons.  Inserts (c) and (d) show the  total widths versus the corresponding masses of the same baryons.
\begin{table}[tr]
\caption{Quantitative information concerning the oscillation behavior of some mesons and baryons analysed previously. P is the period (in MeV), m and w correspond to masses and widths.}
\label{Table I}
\vspace{5.mm}
\begin{tabular}{c c c c c c c c}
\hline\underline{•}
name&q.c.&fig.&J$^{PC}$&mass&P(m)&P(w)&P(m)/P(w)\\
\hline
$f_{0}$&$q{\bar q}$&1&$0^{++}$&475&647&358&1.8\\
$N^{*}$&qqq&5&1/2$^{+}$&939&146.4&352&.42\\
$\Lambda$&qqs&6&1/2$^{+}$&1115.7&169.5&207&.82\\
$\Delta$&qqq&6&3/2$^{+}$& 1232&282.7&276.5&1.0\\ 
$f_{2}$&$q{\bar q}$&1&$2^{++}$&1275&415&358&1.2\\
D&${\it q}{\bar c}$&2&all&1869.6&147&264&.56\\
$D_{S}$&${\it s}{\bar c}$&2&all&1968.5&150.8&157&.96\\
$\Lambda_{C}$&udc&5&1/2$^{+}$&2286.5&377&188.5&2\\
$\Xi$&qss&7&$1/2^{+}$ &1317&377&314&1.2\\
$\Xi_{C}$&qsc&7&$0^{-}$&2467.8&151&157&0.96\\
charm.&$c{\bar c}$&3&all&2981.5&358&308&1.16\\
charm.&$c{\bar c}$&4&$0^{-+}$&2981.5&358&308&1.16\\
botto.&$b{\bar b}$&3&all&9391&408&308&1.32\\
botto.&$b{\bar b}$&4&$0^{-+}$&9391&452&308&1.47\\
\hline
\end{tabular}
\end{table}
\begin{figure}[h]
\hspace*{-0.2cm}
\scalebox{0.58}[0.58]{
\includegraphics[bb=9 134 539 548,clip,scale=0.8] {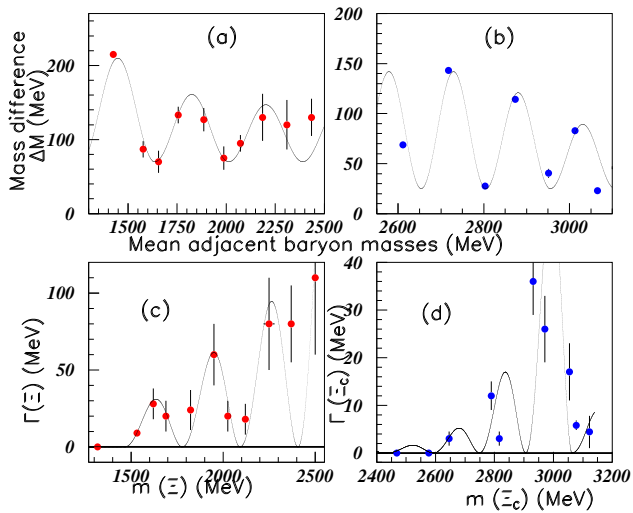}}
\caption{Color on line. Inserts (a) and  (b) show successively the mass difference between successive masses, plotted versus both corresponding mean masses for $\Xi$ and $\Xi_{C}$ baryons.  Inserts (c) and (d) show the  total widths versus the corresponding masses of the same baryons.}
\end{figure}

Fig. 7 shows the results for  $\Xi$ and $\Xi_{C}$ baryons. The fits describe well the data. 

A curious feature is that the extracted periods are also showing regular oscillations. Fig. 8 shows, respectively in inserts (a), (b), and (c) the mass oscillation periods, the width oscillation periods and the ratio of mass over width periods, plotted versus the lower  mass of each family. Full red circles (blue squares) show the results for mesons (baryons) illustrated above. We observe again oscillating shapes. Although we do not have, presently, any motivation to justify the need for observing oscillations in the width periods, nor in the mass over width ratio periods, Fig. 8(b)  and (c) shows that this is however the case. 
\begin{figure}[h]
\hspace*{-0.2cm}
\scalebox{0.58}[0.58]{
\includegraphics[bb=4 80 536 548,clip,scale=0.8] {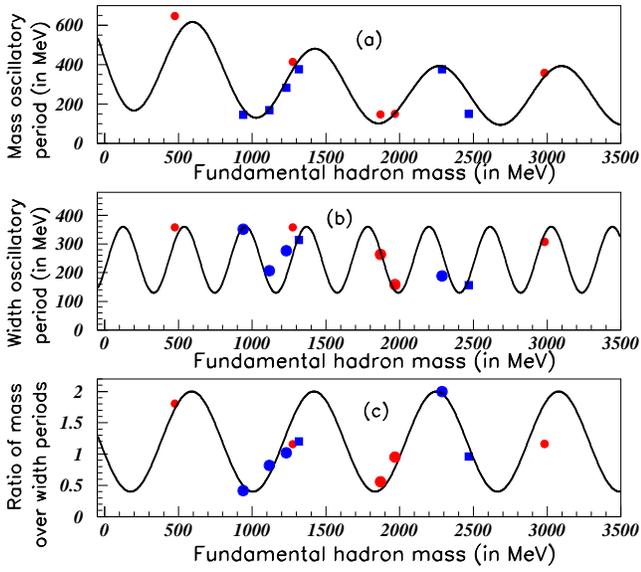}}
\caption{Color on line. Inserts (a), (b), and (c) show successively the variation of the oscillating periods of the hadronic mass families, the corresponding widths, and the ratios of mass periods over the widths periods are plotted versus the lower mass of all hadronic families. Full red circles (blue squares) show the results for mesons (baryons).}
\end{figure}

  The same period P= 829 MeV is observed in inserts (a) and (c). The period of the width period is two times smaller P=415 MeV, otherwise the baryon data, mainly the first one, is badly fitted in insert (b).

 In conclusion, the paper shows that regular oscillations are  often  observed in hadron observables. This property allows to tentatively predict some hadronic masses, still not observed. The same variation of the periods is observed from the analysis of masses and widths of different meson and baryon families. Such behavior requires the need for a theoretical study to describe the observed oscillating distributions and particularly the oscillation amplitudes. Such work is clearly outside the scope of the present work.

\end{document}